\documentstyle[prd,tighten,aps]{revtex}
\begin{document}
\draft

\twocolumn[\hsize\textwidth\columnwidth\hsize\csname
@twocolumnfalse\endcsname
\renewcommand{\theequation}{\thesection . \arabic{equation} }
\title{\bf Perdurance of multiply connected de Sitter space}

\author{Pedro F. Gonz\'alez-D\'{\i}az }
\address{Centro de F\'{\i}sica ``Miguel Catal\'an'',
Instituto de Matem\'aticas y F\'{\i}sica Fundamental,\\
Consejo Superior de Investigaciones Cient\'{\i}ficas,
Serrano 121, 28006 Madrid (SPAIN)}
\date{December 29, 1998}

\maketitle

\begin{abstract}
This paper deals with a study of the effects that spherically symmetric
first-order metric perturbations and vacuum quantum fluctuations
have on the stability of the multiply connected de Sitter
spacetime recently proposed by Gott and Li. It is the main
conclusion of this study that although such a spacetime is stable to
the classical metric perturbations for any size of the
nonchronal region, it is only stable against
the quantum fluctuations of vacuum if the size of the multiply
connected region is of the order the Planck scale.
Therefore, boundary conditions for the state of the
universe based on the notion that the universe created itself
in a regime where closed timelike curves were active and stable,
still appear to be physically and philosophically so well supported
as are those boundary conditions relying on the notion that
the universe was created out of nothing.
\end{abstract}


\vskip2pc]

\renewcommand{\theequation}{\arabic{section}.\arabic{equation}}

\section{\bf Introduction}
\setcounter{equation}{0}

When he was arguing in support of his favourite condition,
Hawking considered [1] that the choice of the boundary conditions
for the quantum state of the universe should be done on just
compact metrics or noncompact metrics which are asymptotic to
metrics of maximal symmetry. While completedness seems to be
an appealing property, spaces which are maximally symmetric are
wanted for their ellegance and greatest simplicity. At the time
being, the quest for the boundary conditions of the universe
has already collected a number of proposals with the above or
similar properties, including the Vilenkin's tunneling wave
function [2], the vanishing Weyl-tensor condition of Penrose
[3] and, perhaps most popularly, the no boundary proposal of
Hartle and Hawking [4]. They have all been much discussed from
different perpectives [5], particularly by invoking their
capability to generate a suitable inflationary mechanism able
to solve original problems of standard cosmology [5,6], giving
at the same time rise to a scale invariant spectrum of density
fluctuations [7,8], compatible with the present observational
status of the universe.

Recent observations on the density of gravitational lenses in
the universe [9,10] and estimates of the present value of the
Hubble constant [11,12] have nevertheless led to the conclusion
that the critical density $\Omega$ has to be smaller than unity,
so favouring an open rather than closed model for the universal
expansion. The problem then is that most of the inflationary
models compatible with the proposed boundary conditions predict
a value for $\Omega$ very close to unity. In order to solve
this problem, Hawking and Turok have recently suggested [13]
the existence of a singular instanton which is able to generate
an open inflationay universe within the framework of the no
boundary proposal. Vilenkin has argued [14] that the singularity
of this instanton may have catastrophic consequences and,
therefore, some procedures have been advanced [15,16] to make
the instanton nonsingular, while still inducing an open inflationary
process.

It is worth noting that so early as 1982 Gott already proposed [17] a
procedure leading to an open inflationary model for the universe.
At the time, of course, it was largely disregarded, since it was
then generally believed that the universe is closed. Motivated by
the present observational status, Gott and Li have now suggested
[18] new boundary conditions based on the Gott's original model
of open inflation. These boundary conditions assume the existence
of a nonchronal region with closed timelike curves (CTC's) in de
Sitter space, separated from the observable universe by a
chronology horizon [19]. The resulting model can be confronted
to e.g. the most popular no boundary paradigm as follows.
Consider first de Sitter space and visualize it by means of the
Schr\"odinger five-hyperboloid with Minkowskian coordinates
[20,21]. If we slice this space along surfaces of constant
timelike coordinate, then the slices become three-spheres and
represent a closed universe. On this slicing both the no
boundary and the Vilenkin conditions hold, so that in these
models the beginning of the universe can be pictured (in the
Euclidean framework) as being the south pole of the Earth
[22], with the time running along the Earth's meridians. Thus,
asking what happened before the beginning of the universe
is like asking what is south of the south pole. Therefore,
these boundary conditions can be regarded to implement the
idea that the universe was created out of nothing. In
contraposition to this approach, one can also slice the
Schr\"odinger's five-hyperboloid vertically, i.e., along
the only spacelike direction defined in terms of the proper
time [18]. The resulting slices are negatively curved surfaces,
describable in terms of open cosmological solutions. It is on this
slicing that the Gott-Li's boundary condition can be defined.
Since the time should now lie along the Earth's Ecuator (or
actually any of its parallels), the origin of time can by no
means be visualized or fixed, so asking what was the earliest
point is like asking what is the easternmost point on the
Earth's Ecuator: there will always be an eastern (that is, earlier)
point. This was implied to mean [18] that on some region of
the slices there must be CTC's. These curves will make the
job of always shifting the origin of the cosmological time.
Saying then that the universe was created from nothing would
be meaningless; what one should instead say is that it created
itself [18].

We have thus two different types of boundary conditions of the
universe that can induce it to be open. Whereas the no boundary
condition does it by rather an indirect way which involves some
suitably modified version of the Hawking-Turok instanton, the
Gott-Li's proposal creates the open universe directly. Is the
latter proposal therefore more fashionable than the former?.
A positive answer to this question could only be made once
the Gott-Li's model would satisfactorily pass some important
tests on its consistency. First of all, one had to check whether
the multiply connected de Sitter space is classically and
quantum-mechanically stable. Li and Gott claimed [23] that
all multiply connected spacetimes with a chronology horizon
(derived from Misner space) are stable to quantum fluctuations
of vacuum, but previous work by Kay, Radzikowski and Wald [24]
and by Cassidy [25] has raised compelling doubts on this conclusion.
The present paper aims at partly filling the above requirements
by studying the classical and quantum stability properties of
the multiply connected de Siiter space. This will be done using
both, a first-order perturbation procedure paralleling the
method devised by Regge and Wheeler [26] to investigate the
stability of Schwarzschild spacetime, and a time-quantization
procedure [27] to analyse regularity of the solution against
vacuum quantum fluctuations. Our main conclusion is that the
multiply connected de Sitter space is stable both classically
and quantum-mechanically. Quantum stability is however restricted
to hold only on the very small regions where the time shows
its essential quantum character.

We outline the paper as follows. In Sec. II we briefly review
how a multiply connected de Sitter space can be constructed, and
show  why its nonchronal region must be confined inside the
cosmological horizon. The stability of the whole de Sitter
space has already been investigated in terms of a global
Friedmann-Robertson-Walker metric [28,29]. However, as far as
I know, no corresponding research has been hitherto attempted
for the de Sitter region covered by static coordinates, that
is the region where multiply connectedness and CTC's should
appear. We have performed this study here, first for the simply
connected static case in Sec. III, and then for the multiply
connected case in Sec. IV. This Section also contains the
analysis of the stability of multiply connected de Sitter space
against quantum fluctuations of vacuum. Finally, we summarize and
conclude in Sec. V.

\section{\bf The multiply connected de Sitter space}
\setcounter{equation}{0}

de Sitter space is usually identified [21] as a maximally symmetric
space of constant negative curvature (positive Ricci scalar) which
is solution to the vacuum Einstein equations with a positive
cosmological constant, $\Lambda>0$. Following Schr\"odinger
[20,29], it can be visualized as a five-hyperboloid defined by
\begin{equation}
w^2+x^2+y^2+z^2-v^2=\rho_0^2 ,
\end{equation}
where $\rho_0=\sqrt{3/\Lambda}$. This hyperboloid is embedded in
$E^5$ and the most general expression for the metric of the
de Sitter space is then that which is induced in this embedding,
i.e.:
\begin{equation}
ds^2=-dv^2+dw^2+dx^2+dy^2+dz^2 ,
\end{equation}
which has topology $R\times S^4$, invariance group $SO(4,1)$
and shows ten Killing vectors (four boosts and six rotations).

Metric (2.2) can be coveniently exhibited in either global or
static coordinates. Global coordinates $t'\in(-\infty,\infty)$,
$\psi_3,\psi_2\in(0,\pi)$, and $\psi_1\in(0,2\pi)$ can be
defined by [29]
\[z=\rho_0\cosh\left(t'/\rho_0\right)\sin\psi_3\sin\psi_2\cos\psi_1 \]
\[y=\rho_0\cosh\left(t'/\rho_0\right)\sin\psi_3\sin\psi_2\sin\psi_1 \]
\begin{equation}
x=\rho_0\cosh\left(t'/\rho_0\right)\sin\psi_3\cos\psi_2
\end{equation}
\[w=\rho_0\cosh\left(t'/\rho_0\right)\cos\psi_3\]
\[v=\rho_0\sinh\left(t'/\rho_0\right). \]
In terms of these coordinates metric (2.2) becomes
\begin{equation}
ds^2=-dt'^2+\rho_0^2\cosh^2\left(t'/\rho_0\right)d\Omega_3^2 ,
\end{equation}
where $d\Omega_3^2$ is the metric on the unit three-sphere. Metric
(2.4) is a $k=+1$ Friedmann-Robertson-Walker metric whose spatial
sections are three-spheres of radius $\rho_0\cosh(t'/\rho_0)$.
Coordinates (2.3) enterely cover the four-dimensional de Sitter
space which would first contract until $t'=0$ and expand
thereafter to infinity.

In order to exhibit metric (2.2) in static coordinates
$t\in(-\infty,\infty)$, $\psi_3,\psi_2\in(0,\pi)$, $\psi_1\in(0,2\pi)$,
one can use the definitions [29],
\[z=\rho_0\sin\psi_3\sin\psi_2\cos\psi_1 \]
\[y=\rho_0\sin\psi_3\sin\psi_2\sin\psi_1 \]
\begin{equation}
x=\rho_0\sin\psi_3\cos\psi_2
\end{equation}
\[w=\rho_0\cos\psi_3\cosh\left(t/\rho_0\right) \]
\[v=\rho_0\cos\psi_3\sinh\left(t/\rho_0\right). \]
Setting $r=\rho_0\sin\psi_3$ (i.e. defining $r\in(0,\rho_0)$),
we obtain the static metric in de Sitter space
\begin{equation}
ds^2=-\left(1-\frac{r^2}{\rho_0^2}\right)dt^2
\left(1-\frac{r^2}{\rho_0^2}\right)^{-1}dr^2+r^2 d\Omega_2^2 ,
\end{equation}
where $d\Omega_2^2$ is the metric on the unit two-sphere.
The coordinates defined by Eqs. (2.5) cover only the portion
of the de Sitter space with $w>0$ and $x^2+y^2+z^2<\rho_0^2$;
i.e., the region inside the particle and event horizons for
an observer moving along $r=0$.

In order to see whether the whole or some restricted region
of the de Sitter space can be made to have multiply
connected topology, with CTC's on it, we will follow the
procedure described by Gott and Li [18], so checking whether
a symmetry like that is satisfied by the Minkowskian covering
to Misner space [21] is somewhere holding in de Sitter space.
On the Minkowskian five-hyperboloid visualizing de Sitter
space, such a symmetry would be expressible by means of the
identification [27]
\[\left(v,w,x,y,z\right)\Leftrightarrow\left(v\cosh(nb)+w\sinh(nb),\right.\]
\begin{equation}
\left.w\cosh(nb)+v\sinh(nb),x,y,z\right),
\end{equation}
where $b$ is a dimensionless arbitrary quantity and $n$ is
any integer number. The boost transformation in the
$(v,w)$-plane implied by this identification will induce a
boost tranformation in the de Sitter space. Hence, since
the boost group in de Sitter space is a subgroup of the de
Sitter group, either the static or the global metric of de
Sitter space can also be invariant under symmetry (2.7).

It is easy to see that there cannot exist any symmetry associated
with identification (2.7) on the $(v,w)$-palne which leaves
metric (2.4) invariant for coordinates (2.3). It follows that
the whole of the de Sitter spacetime can neither be multiply
connected, nor have CTC's. However, for coordinates defined
by Eqs. (2.5) leading to the static metric with an apparent
horizon (2.6), the above symmetry can be satisfied on the region
covered by such a metric, defined by $w>|v|$, where there
are CTC's, with the boundaries at $w=\pm v$ and
$x^2+y^2+z^2=\rho_0^2$ being the Cauchy horizons that limit
the onset of the nonchronal region from the causal exterior
[18]. Such boundaries become then appropriate chronology
horizons for de Sitter space.

\section{Stability of static de Sitter space}
\setcounter{equation}{0}
While our discussion of Sec. II made it clear that multiply
connected de Sitter space is mathematically rich and
interesting, we still need to know if such a space is
indeed a physical object. Therefore, in what follows we shall
use a general-relativity perturbation method to investigate the
stability of the multiply connected de Sitter universe. Since
multiply-connectedness and CTC's only appear in the region
covered by static coordinates, the extension of the analysis
of the cosmologically perturbed global metric for simply
connected de Sitter space [28,29] to a multiply connected
topology would unavoidably lead to rather inconclusive
results. We instead shall proceed as follows. We first extend
the perturbative procedure originally devised by Regge and
Wheeler for the Schwarzschild problem [26] to a cosmological
de Sitter space in this Section, and then in the next Section,
we conveniently include the effects derived from the
identification (2.7) in the resulting formalism. In the present
paper we confine ourselves to the linear analysis, investigating
the stability of simply and multiply connected de Sitter space
in first order perturbation theory by means of
a generalization from the refined
method developed by Vishveshwara [30] and Zerilli [31].

We take as the general background metric $g_{\mu\nu}$ and the
perturbation on it as $h_{\mu\nu}$. The quantity $g_{\mu\nu}$
will be later specialized to be the static de Sitter metric, i.e.,
\begin{equation}
ds^2=-\left(1-H^2 r^2\right)dt^2
+\left(1-H^2 r^2\right)^{-1}dr^2+r^2 d\Omega_2^2 ,
\end{equation}
where we have now denoted $H=\rho_0^{-1}$ for the sake of
simplicity, and $x^0=t$, $x^1=r$, $x^2=\theta$, $x^3=\phi$.
Metric (3.1) corresponds to the initial time-independent
equilibrium configuration, so the problem to be solved is,
then, if metric (3.1) is somehow perturbed, whether the
perturbations will undergo oscillations about the equilibrium
state, or will grow exponentially with time. The static de
Sitter space will be stable in the first case and instable in
the second one.

Since the background is spherically symmetric, any arbitrary
perturbation can be decomposed in normal modes given by [26]
\[\sum f_0(t)f_1(r)f_2(\theta)f_3(\phi) .\]
Associated with these modes we have an angular moment $\ell$
and its projection on the $z$ axis, $M$. For any given value
of $\ell$ there will be two independent classes of perturbations
which are respectively characterized by their parities $(-1)^{\ell}$
(even parity) and $(-1)^{\ell+1}$ (odd parity). Furthermore,
since the background is time-independent, all time dependence
of the perturbations will be given by the simple factor
$\exp(-ikt)$, where $k$ is the frequency of the given mode.

In order to derive the equations governing the perturbations,
we shall start with the Einstein equations
\begin{equation}
R_{\mu\nu}(g)=\Lambda g_{\mu\nu} ,
\end{equation}
with $g$ denoting the de Sitter background metric. For the
perturbed spacetime, these field equations transform into
\begin{equation}
R_{\mu\nu}(g)+\delta R_{\mu\nu}(h)=\Lambda\left(g_{\mu\nu}
+h_{\mu\nu}\right),
\end{equation}
for small perturbation, with $\delta R_{\mu\nu}$ assumed to
contain first-order terms in $h_{\mu\nu}$ only. Now, since
Einstein equations are still valid in the perturbation scheme,
we obtain in this approximation that the differential equations
that govern the perturbations should be derived from the
equations
\begin{equation}
\delta R_{\mu\nu}(h)=\Lambda h_{\mu\nu}.
\end{equation}
The $\delta R_{\mu\nu}$'s will be here computed using the same
general formulas as those employed by Regge and Wheeler [26]
and Eisenhart [32]; i.e.
\begin{equation}
\delta R_{\mu\nu}=-\delta\Gamma_{\mu\nu;\beta}^{\beta}
+\delta\Gamma_{\mu\beta;\nu}^{\beta} ,
\end{equation}
where the semicolon denotes covariant differentiation, and the
variation of the Christoffel symbols is given by
\begin{equation}
\delta\Gamma_{\mu\nu}^{\beta}
=\frac{1}{2}g^{\beta\alpha}\left(h_{\mu\alpha;\nu}
+h_{\nu\alpha;\mu}-h_{\mu\nu;\alpha}\right) .
\end{equation}

After introducing suitable gauge transformations [30,31], the
most general perturbations in static de Sitter space can be
written in forms which are similar to those obtained for
Schwarzschild space; i.e.

For odd parity:

\[h_{\mu\nu}=
\left[
\begin{array}{cccr}
0&0&0&h_0(r)\\
0&0&0&h_1(r)\\
0&0&0&0\\
sym&sym&0&0\\
\end{array}
\right]\]
\begin{equation}
\times\left[\sin\theta\left(\frac{\partial}{\partial\theta}\right)\right]
P_{\ell}(\cos\theta)\exp(-ikt) .
\end{equation}

For even parity:

\[h_{\mu\nu}=\]
\[\left[
\begin{array}{cccr}
H_0\left(1-H^2 r^2\right)&H_1&0&0\\
H_1&H_2\left(1-H^2 r^2\right)^{-1}&0&0\\
0&0&Kr^2&0\\
0&0&0&Kr^2\sin^2\theta\\
\end{array}
\right]\]
\begin{equation}
\times P_{\ell}(\cos\theta)\exp(-ikt) .
\end{equation}
In these expressions "sym" indicates that $h_{\mu\nu}=h_{\nu\mu}$,
$P_{\ell}(\cos\theta)$ is the Legendre polynomial, and $h_0$,
$h_1$, $H_0$, $H_1$, $H_2$ and $K$ are given functions of the
radial coordinate $r$ which must be determined as solutions
to the respective wave equations, subject to suitable boundary
conditions. In order to ensure holding of regularity on the
cosmological horizon at $r=H^{-1}$, we should transform the
components of the perturbations in Eqs. (3.7) and (3.8) to
the representation where the static metric is maximally extended
and regular also on the cosmological horizon. In Kruskal
coordinates $u,v$, the static de Sitter metric can thus be
written as
\begin{equation}
ds^2=F(r)^2\left(du^2-dv^2\right)+r^2 d\Omega_2^2 ,
\end{equation}
where
\begin{equation}
F(r)^2=-\left(\frac{1+Hr}{H}\right)^2
\end{equation}
\begin{equation}
v^2-u^2=\frac{1-Hr}{1+Hr}
\end{equation}
and
\begin{equation}
Ht=-\tanh^{-1}\left(\frac{v}{u}\right) .
\end{equation}

In terms of the Kruskal coordinates, the components of the metric
perturbations take the form (the components that invove only angular
coordinates are the same in the two coordinate systems):
\[h_{00}^K=
\frac{F^2}{u^2-v^2}\left[u^2\left(1-H^2 r^2\right)^{-1}h_{00}\right.\]
\[\left.+v^2\left(1-H^2 r^2\right)h_{11}-2uvh_{01}\right]\]
\[h_{11}^K=
\frac{F^2}{u^2-v^2}\left[v^2\left(1-H^2 r^2\right)^{-1}h_{00}\right.\]
\[\left.+u^2\left(1-H^2 r^2\right)h_{11}-2uvh_{01}\right]\]
\[h_{01}^K=
\frac{F^2}{u^2-v^2}\left[\left(u^2+v^2\right)h_{01}\right.\]
\begin{equation}
\left.-uv\left(\left(1-H^2 r^2\right)^{-1}h_{00}
+\left(1-H^2 r^2\right)h_{11}\right)\right]
\end{equation}
\[h_{03}^K\propto
\frac{1}{u^2-v^2}\left[uh_{03}-v\left(1-H^2 r^2\right)h_{13}\right]\]
\[h_{13}^K\propto
-\frac{1}{u^2-v^2}\left[vh_{03}-u\left(1-H^2 r^2\right)h_{13}\right] , \]
where the superscript $K$ refers to Kruskal coordinates. For
future reference, we introduce here the relation:
\begin{equation}
\exp\left(Hr^{*}\right)=u^2-v^2 ,
\end{equation}
where the new variable $r^{*}$ is defined by
\begin{equation}
r^{*}=H^{-1}\ln\left(\frac{Hr-1}{Hr+1}\right) .
\end{equation}

\subsection{The wave equations for the perturbations}

We derive in what follows the differential equations which should
be satisfied by the perturbations on the maximally extended de
Sitter metric. We shall start with the odd-parity solutions, and
choose the perturbed Einstein equations
\begin{equation}
\frac{kh_0}{1-H^2 r^2}+\frac{d}{dr}\left(1-H^2 r^2\right)h_1=0
\end{equation}
\begin{equation}
\frac{k\left(h_0 '-kh_1-\frac{2h_0}{r}\right)}{\left(1-H^2 r^2\right)}
+(\ell-1)(\ell+2)\frac{h_1}{r^2}=24\pi GH^2 h_1 ,
\end{equation}
where the prime denotes differentiation with respect to radial
coordinate, $^{'}=d/dr$.
Let us first introduce the definition
\begin{equation}
rQ=\left(1-H^2 r^2\right)h_1 ,
\end{equation}
with which Eqs. (3.16) and (3.17) become
\begin{equation}
\frac{kh_0}{\left(1-H^2 r^2\right)}+\frac{d}{dr}(rQ)=0
\end{equation}
\begin{equation}
kh_0 '-k^2 h_1-\frac{2kh_0}{r}+(\ell-1)(\ell+2)\frac{Q}{r}
=24\pi GH^2(rQ) .
\end{equation}
We eliminate then $kh_0$ from Eqs. (3.19) and (3.20) to obtain
\[-\left(1-H^2 r^2\right)\frac{d^2}{dr^2}(rQ)
-\frac{k^2 rQ}{\left(1-H^2 r^2\right)}+\frac{2}{r}\frac{d}{dr}(rQ)\]
\begin{equation}
+(\ell-1)(\ell+2)\frac{Q}{r}=24\pi GH^2(rQ) .
\end{equation}
From Eq. (3.15) we finally obtain the wave equation
\begin{equation}
\frac{d^2 Q}{dr^{*2}}+\frac{1}{4}\left(k^2-V_{eff}\right)Q=0
\end{equation}
where
\begin{equation}
V_{eff}=\left(1-H^2 r^2\right)\left[\frac{\ell(\ell+1)}{r^2}
-24\pi GH^2\right]
\end{equation}
is an effective potential, and
\begin{equation}
h_0=\left(\frac{i}{k}\right)\left(\frac{d}{dr^*}\right)(rQ) .
\end{equation}

The derivation of the wave equation for the case of the even-parity
perturbations has more algebraic complications. From $\delta R_{\mu\nu}
=\Lambda h_{\mu\nu}$ and the general expression (3.8), we obtain first
the independent first-order perturbations of the Ricci tensor
\begin{equation}
\frac{dK}{dr}+\frac{K-h}{r}+\frac{H^2 rK}{\left(1-H^2 r^2\right)}
-\frac{i\ell(\ell+1)h_1}{2kr^2}=0
\end{equation}
\begin{equation}
ikh_1+\left(1-H^2 r^2\right)\left(\frac{dh}{dr}-\frac{dK}{dr}\right)
-2H^2 rh=0
\end{equation}
\begin{equation}
\frac{d}{dr}\left[\left(1-H^2 r^2\right)h_1\right]
+ik(h+K)=0
\end{equation}
\[\left(1-H^2 r^2\right)^2\frac{d^2 h}{dr^2}
+\frac{2}{r}\left(1-H^2 r^2\right)\left(1-3H^2 r^2\right)\frac{dh}{dr}
-k^2 h\]
\[+\frac{2ik\left(1-H^2 r^2\right)}{r^2}\frac{d}{dr}(r^2 h_1)
-\left(1-H^2 r^2\right)\frac{\ell(\ell+1)h}{r^2}\]
\[-4ikH^2 rh_1+2\left(1-H^2 r^2\right)H^2 r\frac{dK}{dr}-2k^2 K\]
\begin{equation}
=-6H^2\left(1-H^2 r^2\right)h
\end{equation}
\[2ik\left(1-H^2 r^2\right)\frac{dh_1}{dr}-2ikH^2 rh_1-k^2 h\]
\[+\frac{\left(1-H^2 r^2\right)^2}{r^2}
\left[\frac{d}{dr}\left(r^2\frac{dh}{dr}\right)
-2\frac{d}{dr}\left(r^2\frac{dK}{dr}\right)\right]\]
\[+\left(1-H^2 r^2\right)\left[\frac{\ell(\ell+1)h}{r^2}
-4H^2 r\frac{dh}{dr}+2H^2 r\frac{dK}{dr}\right]\]
\begin{equation}
=6H^2\left(1-H^2 r^2\right)^2 h
\end{equation}
\[\frac{d}{dr}\left\{\left(1-H^2 r^2\right)\left[\frac{d}{dr}(r^2 K)
-2rh\right]\right\}+\frac{r^2 k^2 K}{\left(1-H^2 r^2\right)}\]
\begin{equation}
-\ell(\ell+1)K-2ikrh_1=-6H^2 r^2 K ,
\end{equation}
where we have used $H_2=H_0\equiv h$, which
is allowed by the high symmetry of the de Sitter space,
and $H_1\equiv h_1$. The three first-order equations (3.25)-(3.27)
can be used to derive any of the subsequent second-order
equations (3.28)-(3.30), provided the following algebraic
relationship is satisfied
\[\left[(\ell-1)(\ell+2)-4H^2 r^2\left(1-3H^2 r^2\right)\right]h\]
\[+\left[-\ell(\ell+1)
+\frac{2\left(k^2 r^2
+1+2H^2 r^2\right)}{\left(1-H^2 r^2\right)}\right]K\]
\begin{equation}
-\left[2ikr+\frac{3iH^2 r\ell(\ell+1)}{k}\right]h_1=0 .
\end{equation}

The wave equation in a single unknown can now be derived from
Eqs. (3.25), (3.26), (3.27) and (3.31). After a lot of algebraic
manipulations we finally obtain
\[\frac{d^2\bar{S}}{dr^{*2}}-\frac{1}{2}\left\{4H^2 r
+\frac{4\left(1-H^2 r^2\right)}{rD(r)}\right.\]
\[\left.\times\left[4H^2 r^2
\left(1-3H^2 r^2\right)-(\ell-1)(\ell+2)\right]\right\}
\frac{d\bar{S}}{dr^*}\]
\[+\frac{1}{4}\left\{k^2+4H^2\left(2+H^2 r^2\right)
-\frac{\left(1-H^2 r^2\right)}{r^2}(\ell-1)\ell+2)\right.\]
\[\left.-\frac{4}{r^2}+\frac{4\left(1-H^2 r^2\right)}{D(r)}
\left[4H^2\left(1-H^2 r^2\right)
\left(1-3H^2 r^2\right)+2k^2\right.\right.\]
\begin{equation}
\left.\left.-\frac{\left(1-H^2 r^2\right)}{r^2}(\ell-1)(\ell+2)
+\frac{3H^2\ell(\ell+1)}{kr}\right]\right\}\bar{S}=0 ,
\end{equation}
in which we have introduced the definition
\begin{equation}
\bar{S}=\left(1-H^2 r^2\right)\frac{h_1}{r} ,
\end{equation}
and the function $D(r)$ has the form
\[D(r)=\frac{2k^2 r^2}{\left(1-H^2 r^2\right)}
-2(\ell-1)(\ell+2)\]
\begin{equation}
+\frac{6H^2 r^2}{\left(1-H^2 r^2\right)}
+4H^2 r^2\left(1-3H^2 r^2\right).
\end{equation}
We note that $D(r)$ approaches infinity as $r$ tends to the
cosmological horizon. The wave equation (3.32) can also be expressed
in terms of another function $S=h_1/r$
\[\frac{d^2 S}{dr^{*2}}-\frac{2\left(1-H^2 r^2\right)}{rD(r)}\]
\[\times\left[4H^2 r^2\left(1-3H^2 r^2\right)
-(\ell-1)(\ell+2)\right]\frac{dS}{dr^*}\]
\[+\frac{1}{4}\left\{3H^2(3+H^2 r^2)
-\frac{\left(1-H^2 r^2\right)}{r^2}(\ell-1)(\ell+2)\right.\]
\[\left.+k^2-\frac{4}{r^2}
+\frac{4\left(1-H^2 r^2\right)}{D(r)}
\left[4H^2\left(1-3H^2 r^2\right)^2+2k^2\right.\right.\]
\begin{equation}
\left.\left.-\frac{\left(1-3H^2 r^2\right)}{r^2}(\ell-1)(\ell+2)
+\frac{3H^2}{kr}\ell(\ell+1)\right]\right\}S=0 ,
\end{equation}
with $D(r)$ also as given by Eq. (3.34), which will be most
useful in what follows.

\subsection{Odd-parity perturbations}

The stability of the static region of de Sitter space to the
odd-parity perturbations will be now examined. As in the
Schwarzschild case [30,31], we shall distinguish two different
cases: we first consider the situation when the frequency $k$
is pure imaginary, and then we analyse the wave problem that
results when that frequency is kept real.

Set $k=i\alpha$. Then, from Eqs. (3.22) and (3.24), we have
\begin{equation}
\frac{d^2 Q}{dr^{*2}}=\frac{1}{4}\left(\alpha^2+V_{eff}\right)Q=0
\end{equation}
\begin{equation}
h_0=\frac{1}{\alpha}\left(\frac{d}{dr^*}\right)(rQ) .
\end{equation}
The coordinate $r^*$ ranges from $-\infty$ to $i\pi/H$. The
upper limit can of course be made zero by a re-definition of
$r^*$, such that now
\[r^*=\frac{1}{H}\ln\left[\frac{1-Hr}{1+Hr}\right] .\]
The effective potential $V_{eff}$ is real and positive
everywhere along this range, and vanishes at $r^*=-\infty$,
i.e., on the boundary at $r=H^{-1}$. The asymptotic solution
to Eq. (3.36) as $r$ approaches the cosmological horizon is
\begin{equation}
Q_H\sim\exp\left(\pm\frac{1}{2}\alpha r^*\right).
\end{equation}
As $r\rightarrow 0$, i.e., as one approaches the other boundary
where we should require the perturbations to fall off to zero,
one can write Eq. (3.36) in the form:
\begin{equation}
\frac{d^2 Q_0}{dr^2}
+\frac{1}{4}\left(24\pi GH^2-\frac{\ell(\ell+1)}{r^2}
-\alpha^2\right)Q_0\simeq 0 ,
\end{equation}
whose general solution can be given in terms of the Bessel
functions [33]
\begin{equation}
Q_0\sim\sqrt{r}{\cal C}_{\frac{1}{2}\sqrt{\ell(\ell+1)+1}}
\left(\frac{1}{2}\sqrt{H^2-\alpha^2}r\right) .
\end{equation}
Since we should require the pertutbations to vanish as
$r\rightarrow 0$, we have to choose for the Bessel function
${\cal C}={\cal J}$ [33]. However, if we take the function
$Q$ to be positive, one can see from Eq. (3.36) that
$d^2 Q/dr^{*2}$ can never become negative within the entire
range of $r$, from 0 to $H^{-1}$, and the solution that goes
to zero at the origin $r=0$ (that is Eq. (3.40) with
${\cal C}={\cal J}$) cannot be matched to the solution that
goes to zero on the horizon at $r=H^{-1}$ (that is Eq. (3.38)
with the sign + in the argument of the exponential.) It follows
that the asymptotic solution near $r=H^{-1}$ has to be
\begin{equation}
Q_H=A\exp\left(-\frac{1}{2}\alpha r^*\right),
\end{equation}
where $A$ is an arbitrary constant, so the radial solution
near $r=H^{-1}$, Eq. (3.37), becomes
\begin{equation}
h_0=-\frac{Q_H}{2H} .
\end{equation}

We can now compute the Kruskal perturbations given by Eqs. (3.13).
Let us concentrate on the expression that results for $h_{03}^K$
near the surface $r=H^{-1}$ (i.e., on $u=v$). Suppressing all
angular dependence, we have
\[h_{03}^K\propto\frac{\left[uh_0-v(1-H^2 r^2)h_1\right]e^{-\alpha t}}
{u^2-v^2}\]
\[=-\frac{\left(\frac{u}{2H}+rv\right)Q_H}{u^2-v^2}\]
\[\simeq -\frac{\left(v+\frac{u}{2}\right)}{H\left(u^2-v^2\right)}
A\exp\left[-\alpha(r^* +t)\right] .\]
Note that the sign of time has been reversed in the above equation
with respect to those perturbations that appear in the
Schwarzschild space. This reversal expresses the fact that the
expansion of the universe is in many ways similar to the collapse
of a star, except that the sense of time is reversed [21].

Using then $\exp(-r^*)=1/(u^2-v^2)^{1/H}$ and $e^t=
\left(\frac{v-u}{u+v}\right)^{1/2H}$, and taking into account
that $u=v$ on the surface $r=H^{-1}$, we finally get
\begin{equation}
h_{03}^K\propto\frac{3}{4}(-1)^{1+\alpha/2H}A
(u-v)^{-\left(1+3\alpha/2H\right)}(u+v)^{-\alpha/2H} .
\end{equation}
Now, since at $t=0$ (i.e. on $u=0$) this perturbation becomes
\begin{equation}
h_{03}^K\propto\frac{3}{4H}(-1)^{-2/H}Av^{-\left(1+2\alpha/H\right)} ,
\end{equation}
which is clearly divergent at $t=0$ on the horizon, and the
physically allowable perturbations should be regular everywhere
in space at $t=0$, we see that this perturbation is physically
unacceptable, and hence cannot exist. It follows that the
odd-parity perturbations with purely imaginary frequency ought
to be ruled out.

Let us consider now the solutions that correspond to real
frequencies $k$. We shall look first at the case of ingoing waves
for which the asymptotic solutions near the horizon at $r=H^{-1}$
are ($V_{eff}=0$):
\begin{equation}
Q_H=A\exp\left(-\frac{ik}{2r^*}\right) ,\;\; r^*\rightarrow-\infty .
\end{equation}
For these solutions it holds
\begin{equation}
h_1=\frac{A\exp\left(-ik/2r^*\right)}{H\left(1-H^2 r^2\right)} ,
\end{equation}
\begin{equation}
h_0=\frac{A}{2H}\exp\left(-ik/2r^*\right) .
\end{equation}
For the cosmological perturbations in Kruskal
coordinates we have then,
\[h_{03}^K\propto\frac{A\left(u/2-v\right)}{H(u^2-v^2)}
\exp\left(-ikr^*/2+ikt\right) .\]
We note that for ingoing waves the horizon $r=H^{-1}$ should
be taken at $u=-v$. Therefore,
\[h_{03}^K\propto\frac{3A}{4H(u+v)}
\exp\left[ik\left(t-r^*/2\right)\right] .\]
Using again $\exp(-r^*)=1/(u^2-v^2)^{1/H}$ and $e^t=
\left(\frac{v-u}{u+v}\right)^{1/2H}$, we finally obtain for
this perturbation
\begin{equation}
h_{03}^K\propto\frac{3A}{4H(u+v)}\left[i(u+v)\right]^{-ik/H} .
\end{equation}

Since $u+v\rightarrow 0$ on the horizon, at first glance this
perturbation appears to be seriously divergent. However, as it
also happens in the Schwarzschild case [30], one can build wave
packets which are convergent
everywhere in space out of the monochromatic waves. If we form
the purely ingoing waves into a wave packet by taking $A$ to be
a function of $k$ given by the Fourier transform of a function
$f(q)=\int A(k)\exp(-ikq)dk$, which vanishes for $q<1$, by
integrating over $k$, Eq. (3.48) transforms into
\begin{equation}
h_{03}^K\propto\frac{3}{4H(u+v)}
f\left\{\frac{1}{H}\ln\left[i(u+v)\right]\right\} .
\end{equation}
There cannot be any singularity from the $(u+v)^{-1}$ factor
in Eq. (3.49) because $f$ is nonzero only when $i(u+v)>1$.
Thus, $h_{03}^K$ does not diverge anywhere in space, but it
is always pure imaginary.

For outgoing waves, the asymptotic solutions near $r=H^{-1}$
are given by ($u=v$):
\begin{equation}
Q_H=A\exp\left(+ik/2r^*\right),\;\; r^*\rightarrow-\infty ,
\end{equation}
and, for wave packets formed as before, we finally obtain
\begin{equation}
h_{03}^K\propto-\frac{3}{4H(u-v)}
f\left\{\frac{1}{H}\ln\left[i(u-v)\right]\right\} ,
\end{equation}
which, although coverging everywhere in space, is always pure
imaginary, such as it happened for perturbation (3.49). The
reason for these perturbations to be imaginary resides in the
fact that the argument of the logarithm in the above expressions
should be larger than unity, which in turn requires that both
$u$ and $v$ are imaginary simultaneously. Note that this does
not imply rotation of the time $t$ into the imaginary axis
because Kruskal coordinates $u$ and $v$ appear in the form
$v/u$ in Eq. (3.12).

Thus, the main conclusion for odd-parity perturbations is that,
either they cannot exist whenever their frequency is purely
imaginary, or they are stable and physically acceptable if
their frequency is real. Since in the latter case the solutions
are purely imaginary, one can also conclude that such
perturbations would not be observable.

\subsection{Even-parity perturbations}

In terms of the most covenient function $S=\bar{S}/(1-H^2 r^2)$,
the second-order wave equation for the even-parity
perturbations given by Eq. (3.35), near the horizon at
$r=H^{-1}$, becomes
\begin{equation}
\frac{d^2 S}{dr^{*2}}+\frac{1}{4}\left(k^2+8H^2\right)S=0 .
\end{equation}
Thus, the asymptotic form of the general solution at $r=H^{-1}$,
$S_H$, would read:
\begin{equation}
S_H\sim\exp\left(\pm i\sqrt{\frac{1}{4}k^2+2H^2}r^*\right) .
\end{equation}
As for the boundary at $r=0$, one attains that no perturbation
can consistently be expected, since from Eq. (3.35) we obtain
$S_0=0$ (unless for $\ell=0$ for which case $S_0$ is an
arbitrary constant), and therefore $h=h_1=0$. Hence, we
can readily get an expression for $h_{00}^K$, independently
of the value of the frequency. It follows that the two signs
involved in the exponent of Eq. (3.53) are allowed and should
therefore be taken into account in our analysis.

Again we first consider the case where the frequency $k$ is
purely imaginary, $k=i\alpha$. Then
\begin{equation}
S_H\sim\exp\left(\pm i\sqrt{2H^2-\frac{1}{4}\alpha^2}r^*\right).
\end{equation}
We have two distinct cases: case I, for which $|\alpha|>\alpha_c
\equiv 2\sqrt{2}H$, and case II, for which $|\alpha|<\alpha_c$.
In case I the asymptotic solution reduces to
\begin{equation}
S_H\sim\exp\left(\pm\xi r^*\right) ,
\end{equation}
where
\begin{equation}
\xi=\sqrt{\frac{1}{4}\alpha^2-2H^2}
\end{equation}
is real. Because case II is qualitatively the same as that
for real frequency (to be dealt with later on),
we shall concentrate now on case I only. Since $S=h_1/r$,
near $r=H^{-1}$, one can assume the asymptotic forms
\[h_1=A\exp\left(\pm\xi r^*\right) ,\;\;
h=B\exp\left(\pm\xi r^*\right) .\]
Choosing first the minus sign in the exponent of these two
functions, from the equation relating radial functions $h$
and $h_1$ to one another (which can be obtained by suitably
combining Eqs. (3.26) and (3.27)), after specializing to
$r=H^{-1}$ and $k=i\alpha$, i.e.
\[4H\frac{dh_1}{dr^*}+4\frac{d^2 h_1}{dr^{*2}}+\alpha^2 h_1
+2\alpha\left(Hh+2\frac{dh}{dr^*}\right)=0, \]
we get
\begin{equation}
B=\frac{2H\xi+4H^2-\alpha^2}{\alpha\left(2\xi-H\right)}A .
\end{equation}
Let us now denote $\alpha=\pm\epsilon H$, with $\epsilon
\geq 2\sqrt{2}$. Then the analysis of all possible resulting
cases (including the use of both signs in the exponent of
Eq. (3.54)) will require considering
\[\xi=\pm\sqrt{\frac{\epsilon^2}{4}-2}H .\]
We shall look at three significant values of $\epsilon$,
namely, $2\sqrt{2}$, 3 and $\infty$, in the following cases:
(i) $\alpha>0$, $\xi>0$ ($B=\sqrt{2}A$ for $\epsilon=2\sqrt{2}$,
$A=0$ for $\epsilon=3$, and $B=-A$ for $\epsilon\rightarrow\infty$),
(ii) $\alpha>0$, $\xi<0$ ($B=\sqrt{2}A$ for $\epsilon=2\sqrt{2}$,
and $A=B$ for $\epsilon=3$ and $\xi\rightarrow\infty$), (iii)
$\alpha<0$, $\xi<0$ ($B=-\sqrt{2}A$ for $\epsilon=2\sqrt{2}$,
$B=-A$ for $\epsilon=3$ and $\epsilon\rightarrow\infty$), and
(iv) $\alpha<0$, $\xi>0$ ($B=-\sqrt{2}A$ for $\epsilon=2\sqrt{2}$,
$A=0$ for $\epsilon=3$, and $A=B$ for $\epsilon\rightarrow\infty$).
Clearly, if all of these particular cases
led to stability of de Sitter space
against the considered perturbations, one could conclude that de
Sitter space is stable to such perturbations in all cases. Thus,
for case I, the relation between the coefficients $A$ and $B$
for the asymptotic solutions $h$ and $h_1$ must run between the
extreme values for $A$ ($B$ fixed), $A=B$ and $A=-B$, passing on
$A=0$.

For $A=B$, the perturbation in the Kruskal coordinates, e.g. $h_{00}^K$,
is given by (angular dependence suppressed)
\[h_{00}^K=\frac{F^2(u-v)^2}{u^2-v^2}he^{-\alpha t}\]
\begin{equation}
=(-1)^{-\alpha/2H}F^2 A
\frac{(u-v)^{1-\sqrt{\alpha^2/4H^2-2}-\alpha/2H}}
{(u+v)^{1+\sqrt{\alpha^2/4H^2-2}+\alpha/2H}} .
\end{equation}
At $t=0$ ($u=0$),
\begin{equation}
h_{00}^K=(-1)^{{1-\sqrt{\alpha^2/4H^2-2}-\alpha/H}}
v^{-\left(2\sqrt{\alpha^2/4H^2-2}+\alpha/H\right)} .
\end{equation}
Eq. (3.59) is divergent as $v\rightarrow 0$, except for the
case $\alpha=-\epsilon H$, $\xi>0$ or $\xi$ very large, for
which case $h_{00}^K=-F^2 A$. Hence, except for this case,
all perturbations are physically unacceptable, as they all diverge
at the initial time $t=0$.

For $B=-A$, we have
\begin{equation}
h_{00}^K=(-1)^{-\alpha/2H}F^2 A
\frac{(u+v)^{1-\sqrt{\alpha^2/4H^2-2}+\alpha/2H}}
{(u-v)^{1+\sqrt{\alpha^2/4H^2-2}+\alpha/2H}} ,
\end{equation}
which at $t=0$ ($u=0$) reduces to
\begin{equation}
h_{00}^K=(-1)^{-\left(1+\sqrt{\alpha^2/4H^2-2}+\alpha/H\right)}
F^2 A v^{-2\sqrt{\alpha^2/4H^2-2}} .
\end{equation}
We note that all of these perturbations are physically
unacceptable, except for the case $\alpha=-\epsilon H$,
$\xi<0$, with $\epsilon=3$, where $h_{00}^K=iF^2 Av$.

Finally, when $A=0$, we obtain
\[h_{00}^K\]
\begin{equation}
=(-1)^{-\alpha/2H}F^2 B(u^2+v^2)
\frac{(u+v)^{\alpha/2H-1-\sqrt{\alpha^2/4H^2-2}}}
{(u-v)^{1+\sqrt{\alpha^2/4H^2-2}+\alpha/2H}} .
\end{equation}
Again at $t=0$ ($u=0$), this perturbation reduces to
\begin{equation}
h_{00}^K=(-1)^{-\left(1+\sqrt{\alpha^2/4H^2-2}+\alpha/H\right)}
F^2 B v^{-2\sqrt{\alpha^2/4H^2-2}} .
\end{equation}
It can easily be checked that in all the cases, without any
exception, Eq. (3.63) diverges as $v\rightarrow 0$, and
therefore this perturbation is physically unacceptable
and should be ruled out.

We are in this way left with two physically acceptable
even-parity perturbations for purely imaginary frequency
in case I: that given by Eq. (3.58) for negative $\alpha$,
positive $\xi$ and very large $\epsilon$, and that given
by Eq. (3.60) for negative $\alpha$, negative $\xi$ and
$\epsilon$ about 3. Since negatives values of $\alpha$
correspond to the case of outgoing perturbations for which
$u=v$ on the horizon $r=H^{-1}$, these perturbations will
be stable as the resulting powers to the factor $(u-v)$
are positive definite in both cases. Note, furthermore,
that at least the perturbation given by Eq. (3.60) is
always purely imaginary. Thus, even-parity perturbations
with purely imaginary frequency in case I either are
physically unacceptable or, unlike in the Schwarzschild space [30],
are stable and most of them imaginary.

In the case that $k$ is kept real the asymptotic general
solution on the cosmological horizon has already been given by
Eq. (3.53). The analysis to follow will also be valid for
purely imaginary frequency satisfying the condition implied
by case II, with the asymptotic solution being given by
Eq. (3.54) in this case. For real frequency, the relation
between $h$ and $h_1$ is
\begin{equation}
i\left[k^2 h_1-4\left(H\frac{dh_1}{dr^*}
+\frac{d^2 h_1}{dr^{*2}}\right)\right]
=2k\left(Hh+2\frac{dh}{dr^*}\right) ,
\end{equation}
where we have specialized to the region $r=H^{-1}$. Let us
now assume, like it was made for purely imaginary
frequencies, that the asymptotic forms for $h$ and $h_1$
near the cosmological horizon are given by
\begin{equation}
h_1=A\exp(-i\nu r^*) ,\;\; h=B\exp(-i\nu r^*) ,
\end{equation}
in which we have introduced the short-hand notation
\begin{equation}
\nu=\sqrt{2H^2+\frac{1}{4}k^2} .
\end{equation}
From Eqs. (3.64) and (3.65) we get
\begin{equation}
B=
-\left[\frac{2\nu\left(k^2+5H^2\right)}{k\left(k^2+9H^2\right)}
+\frac{4iH^3}{k\left(k^2+9H^2\right)}\right]A .
\end{equation}
Restricting to the case $k^2>>H^2$, so that $\nu\simeq k/2$,
we see that there are two solutions: when $k\simeq 2\nu$
(ingoing waves), $A=-B$, and when $k\simeq -2\nu$ (outgoing
waves), $A=B$. Suppressing again the angular dependence of
the perturbations, we can then compute such perturbations in
Kruskal coordinates. In the case of ingoing waves, $u=-v$ on
the horizon, and we have
\begin{equation}
h_{00}^K=F^2 A(-1)^{ik/2H}\frac{(u+v)^{1-ik/H}}{u-v} ,
\end{equation}
and for outgoing waves ($u=v$),
\begin{equation}
h_{00}^K=F^2 A(-1)^{ik/2H}\frac{(u-v)^{1+ik/H}}{u+v} .
\end{equation}
Clearly, the perturbations (3.68) and (3.69) are stable near
the cosmological horizon and everywhere inside it, even in
the forms given by these equations, without building suitable
wave packets by superposing them. This analysis can readily
be generalized to any values of $k$ and $H$, obtaining the
same conclusion.

\section{Stability of multiply connected de Sitter space}
\setcounter{equation}{0}

In Sec. III we have investigated the stability properties
of the simply connected de Sitter region covered by static
coordinates. We had to do so because, as far as I can know,
such an analysis had not been carried out so far, and we
needed it to prepare our system to study the perturbations
when the coordinates involved are identified in such a way
that this region of de Sitter space becomes multiply connected
topologically, such as it was discussed in Sec. II. We have
obtained that on the static region, the de Sitter is also
stable to the first-order perturbations that satisfy its
symmetries.

\subsection{Classical perturbations}

We shall now study the effect that topological
multiple connectedness has on the stability of de Sitter space.
Because of the high symmetry of this space, the time parameter
$t$ always appears in the form of a factorized exponential
factor in all the perturbations, either as $\exp(\pm\alpha t)$,
if the frequency of the perturbative modes is pure imaginary,
or as $\exp(\pm ikt)$, if that frequency is kept real. Thus,
we can generically write the time factor as
$\exp\left[g(k)t\right]$, with $g(k)$ a given function of the
mode frequency. From our discussion in Sec. II, it follows
that de Sitter space can be made multiply connected by simply
including the time identification $t\leftrightarrow t+nb/H$,
with $b$ a dimensionless arbitrary period and $n$ any integer
number. This will amount to the insertion of an additional
factor
\[\exp\left[nbg(k)/H\right] \]
for each value of the integer $n$ in the distinct expressions
for the first-order perturbations obtained in Sec. III. In
order to take into account all possible values of $n$, along
its infinite range, one then should coveniently sum over all
$n$, from 0 to $\infty$, introducing the statistical factor
$1/n!$ to account for the equivalent statistical weight one
must attribute to all of these contributions. Thus, the
general expression for the perturbations in multiply connected
de Sitter space would be
\begin{equation}
h_{ij}^K(b)=
\sum_{n=0}^{\infty}\frac{h_{ij}^K\exp\left[nbg(k)/H\right]}
{n!} ,
\end{equation}
where $h_{ij}^K$ generically denotes the first-order perturbations
in Kruskal coordinates for simply connected de Sitter space
which were computed in Sec. III.

In what follows we shall perform the calculation of the relevant
$h_{ij}^K(b)$ for all physically acceptable perturbations. We
shall first restrict ourselves to the regime where both
$kb/H$ and $\alpha b/H$ are much smaller than unity; i.e.
we will work in the regime characterized by nonchronal
regions and CTC's whose size is very small. As it will be
seen below, this is the regime of most physical interest
where vacuum quantum fluctuations can be kept convergent
everywhere. Let us start with odd-parity perturbations
with real frequency. For the case of ingoing waves, we have
for the asymptotic solution at $r=H^{-1}$ [34]
\[h_{03}^K(b)=
\frac{3A}{4H(u+v)}\left[i(u+v)\right]^{-ik/H}
\sum_{n=0}^{\infty}\frac{e^{inkb/H}}{n!}\]
\begin{equation}
=\frac{3A}{4H(u+v)}\exp\left(e^{ikb/H}\right)
e^{-ik\ln\left[i(u+v)\right]/H} .
\end{equation}
For small values of $kb/H$, Eq. (4.2) can be approximated to:
\begin{equation}
h_{03}^K(b)\simeq
\frac{3Ae}{4H(u+v)}\exp\left\{-\frac{ik}{H}
\left[\ln\left(i(u+v)\right)-b\right]\right\} .
\end{equation}
Forming again a wave packet out of monochromatic perturbations
(4.3), we finally obtain for this type of waves
\begin{equation}
h_{03}^K(b)\simeq
\frac{3e}{4H(u+v)}f\left[\frac{1}{H}\left(\ln(u+v)+\frac{i\pi}{2}
-b\right)\right] ,
\end{equation}
which still is a convergent expression for all times. If we let
$b$ to be complex, so that $b=b+i\pi/2$, then Eq. (4.4)
becomes not only convergent but pure real as well.

For outgoing waves, an analogous calculation leads finally to
the perturbation:
\begin{equation}
h_{03}^K(b)\simeq
-\frac{3e}{4H(u-v)}f\left[\frac{1}{H}\left(\ln(u-v)+\frac{i\pi}{2}
+b\right)\right] ,
\end{equation}
which is also always covergent and pure real if we similarly
let $b$ be complex and given by $b=b-i\pi/2$. We can then
conclude that in the considered regime, multiply connected
de Sitter space is stable to all odd-parity perturbations
that are physically acceptable.

For even-parity perturbations which also are physically
acceptable we have [34]
\begin{equation}
h_{00}^K(b)=
h_{00}^K\sum_{n=0}^{\infty}\frac{e^{-nb\alpha/H}}{n!}
=h_{00}^K\exp\left[e^{-b\alpha/H}\right] ,
\end{equation}
for any value of $kb/H$.
Thus, making the de Sitter space multiply connected preserves
the stability of these perturbations and increases their
amplitude, specially for small values of $b\alpha/H$.

Next we consider even-parity perturbations with real
frequency. We first note that in this case the
perturbations corresponding to the asymptotic
solutions near $r=H^{-1}$ can also be expressed in terms
of wave packets in the simply connected case. They are:
\begin{equation}
h_{00}^K=
F^2\left(\frac{u+v}{u-v}\right)f\left[\frac{1}{H}\left(\ln(u+v)
+\frac{i\pi}{2}\right)\right] ,
\end{equation}
for ingoing waves, and
\begin{equation}
h_{00}^K=
F^2\left(\frac{u-v}{u+v}\right)f\left[\frac{1}{H}\left(\ln(u-v)
+\frac{i\pi}{2}\right)\right] ,
\end{equation}
for outgoing waves. Because of the form of the
Kruskal-coordinate dependent prefactor, these expressions
are real in any case.

When we multiply connect the de Sitter space in the regime
of small values of $kb/H$, Eqs. (4.7) and (4.8) transform
into:
\begin{equation}
h_{00}^K(b)=
eF^2\left(\frac{u+v}{u-v}\right)f\left[\frac{1}{H}\left(\ln(u+v)
+\frac{i\pi}{2}-b\right)\right] ,
\end{equation}
for ingoing waves, and
\begin{equation}
h_{00}^K(b)=
eF^2\left(\frac{u-v}{u+v}\right)f\left[\frac{1}{H}\left(\ln(u-v)
+\frac{i\pi}{2}+b\right)\right] ,
\end{equation}
for outgoing waves. Note that the argument of the function $f$
becomes real when we allow $b$ to be complex and given by
$b=b\pm i\pi/2$, with the + sign for ingoing waves and the
- sign for outgoing waves. Anyway, the perturbations given
by Eqs. (4.9) and (4.10) keep being convergent and, therefore,
one can conclude that in the regime of very small values of
$b/H$, the multiply connected de Sitter space is also stable
to first-order perturbations which respect the symmetry of
this space.

As to the perturbations for larger values of $b/H$ and real
frequency, we first note that the expressions for the
components $h_{03}^K(b)$ and $h_{00}^K(b)$, before forming
any wave packets, are given by expressions which are the
same as those obtained above in the regime of very small
values of $b/H$, but with the parameter $b$ replaced for
the $k$-dependent function $-\frac{iH}{k}\exp\left(ikb/H\right)$.
The wave packets formed using the same procedure as for all
the above cases will therefore involve characteristic
functions of the form
\[f\left\{\frac{H}{b}\int_{0}^{1}
\frac{dxbe^x}{b+\ln\left[i(u\pm v)\right]}\right\} ,\]
where $x=\exp(ikb/H)$, and the sign +/- in the argument of the
logarithm stands for ingoing/outgoing waves. Now, for
$b>>\ln\left[i(u\pm v)\right]$ we can obtain [33,34]
\[f\left[\frac{H}{b}Ei(x)|_{0}^{1}\right] ,\]
with $Ei(x)$ the exponential integral function. Since the
argument of $f$ is then always smaller than unity, we have
$f=0$ on this regime.

As $(u\pm v)$ becomes very small, so that $b<<y=|{\rm Re}
\ln\left[i(u\pm v)\right]|$, the wave-packet function
approaches the form [33,34]
\[f\left\{\frac{H}{b}\lim_{y\rightarrow\infty}
\left[\frac{\Phi(y,y+1;1)}{y}\right]\right\}=0 ,\]
with $\Phi$ the degenerate (confluent) hypergeometric
function. Thus, also for arbitrarily large nonchronal regions,
the multiply connected de Sitter space is stable to all
physically allowable classical perturbations.

\subsection{Quantum fluctuations}

In what follows we shall briefly discuss the possible influence
that multiply connectedness may have on the quantum stability
of the de Sitter space. Because of the presence of a chronology
horizon on the surface $r=H^{-1}$, it could at first sight be
thought that the quantum renormalized stress-energy tensor,
$\langle T_{\mu\nu}\rangle_{ren}$, for vacuum quantum
fluctuations generated in multiply connected de Sitter space
ought to diverge [35]. However, it has recently been stressed
that this could not be actually the case if either we
consistently redefine the quantum vacuum [23], or we introduce
a suitable quantization of the relevant time parameter, beyond
semiclassical approximation [27]. To see how these ideas apply
to the case under study, let us work in the Euclidean
framework where the Kruskal metric is obtained by rotating
coordinates $u$ and $t$ to the imaginary axis, starting with
Eq. (3.9); i.e., $u=i\eta$ and $t=i\tau$. We get then
\begin{equation}
ds^2=\frac{4}{H^2(1+\eta^2+v^2)}\left(d\eta^2+dv^2\right)
+r^2 d\Omega_2^2 ,
\end{equation}
which in fact is definite positive. Corresponding to this
metric, Euclidean time $\tau$ will be defined by the relation
\begin{equation}
\exp(2iH\tau)=\frac{v-i\eta}{v+i\eta} .
\end{equation}
Wick rotating also in the identification $t\leftrightarrow
n\beta$ (where the nonperiodic time term is disregarded and
$\beta=b/H$ is the period) that makes de Sitter space
multiply connected, Eq. (4.12) transforms into
\begin{equation}
\exp(2inH\beta)=\frac{v-i\eta}{v+i\eta} ,
\end{equation}
from which one can get the complex relation
\begin{equation}
v-i\eta=\sqrt{v^2+\eta^2}\exp(inH\beta) .
\end{equation}
It follows from Eq. (4.14) that the Euclidean time preserves a
periodic character also on the Euclidean sector: $\beta=2\pi/H$,
that is to say, $b=2\pi$. This result can be interpreted along
the following lines. First of all, one can readily see that
multiply connectedness in de Sitter space is nothing but the
Lorentzian counterpart of the existing thermal states that are
uncovered in the Euclidean description [36]. This relation might
be reflecting the origin of the excess of some perturbative
waves which we have found above for multiply connected de Sitter
space with respect to its simply connected space.

Quite more importantly, the value $b=2\pi$ can be used (as
Gott and Li did [23]) to redefine a conformal vacuum in Euclidean
space for which $\langle T_{\mu\nu}\rangle_{ren}$ does not
diverge even on the chronology horizon. However, the meaning of
this horizon in such a vacuum has been discussed by Kay,
Radzikowski and Wald [24] and Cassidy [25], so that some quite
founded doubts can be cast on its real existence and capability to
restore quantum stability this way. But if we adhere to the also
recently suggested [27] kind of time quantization by which
$t=(n+\gamma)t_0$ (with $t_0$ a constant time whose value is of
the order the Planck time, and
$\gamma$ the automorphic parameter [37,38], $0\leq\gamma\leq 1/2$),
and note the formal analogy of this expression with that is implied
by the identification $t\leftrightarrow t+nb/H$ when we take $b=2\pi$
and $t=2\pi\gamma/H$, we see that quantum stability could be
unambiguously restored in multiply connected de Sitter space,
provided we accept restricting the nonchronal region and the
CTC's on it to be essentially at the Planck scale, $1/H\sim l_{p}$
[27].

\section{Summary and conclusions}

The main aim of this paper is to study the classical and quantum
stability of the multiply connected de Sitter space. This space
arises when we introduce some periodicity conditions on the
coordinates describing the five-dimensional Minkowski hyperboloid
and can only be exhibited on the region covered by the static
de Sitter coordinates. Using a first-order perturbation
formalism analogous to that which was originally developed by
Regge and Wheeler for Schwarzschild metric [26], we have shown
that multiply connected de Sitter space is classically stable
to these perturbations, no matter the size of the static region.
Although stability aginst higher-order perturbations has not been
checked in this paper, one would expect these perturbations not
to introduce any instabilities, such as it occurs in
Schwarzschild spacetime [26].

By continuing the Kruskal extension of the multiply connected,
static de Sitter metric into its Euclidean section, we have
also argued that quantum vacuum fluctuations should not
induce any divergence on this space, provided the nonchronal
region and the CTC's on it are all sufficiently small, probably
of the order the Planck size [27]. We therefore consider multiply
connected de Sitter universes to be genuine components of any
future description of a well-defined theory of quantum gravity.
In particular, the considered stable little multiply connected
universes should be included, together with Euclidean and
multiply connected wormholes [19], ringholes [39], Klein
bottleholes [40] and virtual black holes [41], as components
of the vacuum quantum spacetime foam, where their CTC's
would contribute the required violation of causality that
governs the foam. Thus, Planck-sized de Sitter universes containing
CTC's can help to define the boundary conditions of the universe
we live in, probably along the lines recently suggested by
Gott and Li [18] and discussed in the Introduction of this
paper.

To close up, I would like to refer to the interesting possibility
that the Hartle-Hawking and the Gott-li conditions might both
be seen to imply initial physical pictures which, at least in
a way, appear to be complementary. In classical relativity
time and spatial coordinates can still be distinguished by
the fact that, whereas spatial coordinates do not single out
a particular sign to run over, time can only run forwards.
This does not help, none the less, to understand the Lorentzian
signature of the classical metrics which are not positive
definite. However, as we approach the regime of the quantum
spacetime foam, all of this distinction can be thought to
vanish, since in such a regime there will be CTC's
everywhere and hence the two time direction would become
equally allowable and, at the same time, metrics can be taken
to be positive definite. In order to describe the quantum
origin of the universe, one then can either keep CTC's and
Lorentzian signatures simultaneously, as assumed by Gott
and Li, or disregard CTC's while using Euclidean signature
where time becomes spacelike, as suggested by Hartle and
Hawking. As seen in this way, the two pictures would actually
describe rather complementary aspects of the initial physical
situation.

\acknowledgements

\noindent  For hospitality at DAMTP, where part of this work was done,
the author thanks S.W. Hawking. This reasearch was supported
by DGICYT under
Research Project No. PB97-1218.

\end{document}